\documentclass[12pt]{article}
\usepackage{float}
\usepackage{amsmath}
\usepackage{amsfonts}
\usepackage{amssymb}
\usepackage{graphicx}
\usepackage{cite}
\usepackage{color}
\usepackage{dcolumn}
\usepackage{wrapfig}
\usepackage{longtable}
\usepackage{times}
\usepackage{caption}
\usepackage{subcaption}
\usepackage{bm}
\usepackage[margin=0.7in]{geometry}
\usepackage{newfloat}
\DeclareFloatingEnvironment[name={SM Figure}]{suppfigure}
\DeclareFloatingEnvironment[name={SM Table}]{supptable}
\begin{document}
\begin{flushleft}
{\LARGE{\bf Nano radiation detector: A possibility}}
\\
\vspace{0.1cm}
{\large {\bf Making Physically Useful Nanostructures With Focused Ion Beam: A study of Nano Gaps}}
\\
%{\small{\bf Masters Thesis Report }}

\bf Saptarshi Ghosh$^{1\dagger}$ and H.C.Verma$^{2\ast}$
\\
\vspace{0.2cm}
\it ${^1}$ Discipline of Physics, Indian Institute of Technology Indore, Indore 453552, India
\\
\it ${^2}$ Department of Physics, Indian Institute of Technology, Kanpur 208016, India
\\
${^{\dagger}}$ Email: sapta15@gmail.com
${^{\ast}}$ Email: hcverma@iitk.ac.in
\end{flushleft}

\begin{abstract}
	Child-Langmuir law expresses the IV relation between two parallel plates. It has been found that in reality the ordinary Child Langmuir Law does not hold true. It shows some geometric dependence on the system.In this report capacitors of various thickness and septation has been studied both in macro and nano scale to propose a geometrical correction to the law. More over from the studies, a new type of radiation detector was prepared which can detect radiation in small volumes as compared with normal radiation detectors.
\end{abstract}

\noindent {\bf keywords:} Child-Langmuir law, Nano-Capacitors

\section{Introduction} 
In modern days physicists are trying to study local
effects inside different types of materials. These become effective in Nano meter and Micro meter range. So the importance of studying the Nano structures and designing them for different purposes is of immense importance.With increasing use of Nano/Micro scale structures, in
scientific researches and in industrial applications, the importance of the field of Nano Science has increased.The increasing use of Micro structures in present day circuitry of different appliances has reduced their size dramatically and has given them more compact look.For this reason, the interest now lies in further reducing
the size of the materials and that makes Nano Technology and Nano Science to be one of the forefront of active
research. \\
Nuclear radiations $\alpha, \beta, \gamma $ particles have been used for a variety of material modifications. Detecting their presence is therefore quite important task. The present day detectors are capable of detecting such radiations in a large volume. With circuitry shrinking to nano and micro scale , it will be very useful to have particle radiation detectors with smaller dimensions. The present work is a step in this direction.\\ 
The report is divided in five sections. Firstly,The project framework and the main motivation behind the studies has been discussed.Then the studies on Macro and Nano scale is presented along with the fabrication techniques and experiment details. Later the Nano Radiation Detector (NRD) model is presented along with the experimental data.At the end Further improvement upon the present NRD model has been suggested.
\section{Project Framework}
The main objective of the project is the nano radiation detector.Various studies have been made for detector designs. In last semester, due to technical problems macro capacitors and nano capacitors could not be studied.In section 3 of the report, these objectives were fulfilled.
The proposed detector is based on passes of current through the medium in a small size capacitor and hence IV characteristic of such a capacitor has to be studied. Though Child-Langmuir Law gives some insight but the theoretical factors are to be studied more systematically. We have done some work in the direction and the results are presented in this report.\\
The effects of $\alpha$ Particles on IV characteristic is also reported and this shows the feasibility of out setup to be developed as $\alpha$ particle detectors.
\section{Geometry effect on I-V characteristic of a gap between parallel plate capacitor}
\begin{figure}[htbp]
	\begin{center}
		\includegraphics[width=0.4\textwidth]{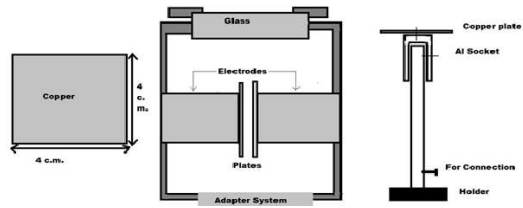}
		\caption{Vacuum Chamber With parallel plate Capacitor}
		\label{fig:chamber}
	\end{center}
\end{figure}

\begin{figure}[htbp]
	\begin{center}
		\includegraphics[width=0.2\textwidth]{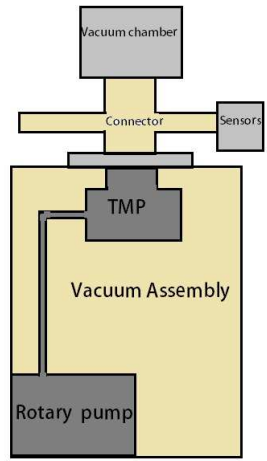}
		\caption{TMP Vacuum Assembly}
		\label{fig:system}
	\end{center}
\end{figure}

\begin{figure}[htbp]
	\begin{center}
		\includegraphics[width=0.2\textwidth]{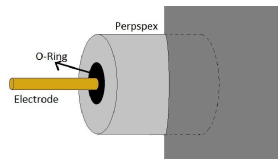}
		\caption{Vacuum Chamber With parallel plate Capacitor}
		\label{fig:electrode}
	\end{center}
\end{figure}

\paragraph{\normalsize{\centerline{\textbf{3.1 Studies on Macro capacitor}}}} \hspace{1pt} \vspace{0pt} \\
\subparagraph{3.1.1 Fabrication} \hspace{1pt} \vspace{0pt} \\
The system used to perform the study mainly consists of a vacuum chamber and a Turbo Molecular Pump (TMP) assembly. For studying the geometric effect both the chamber and the assembly were designed from scratch.\\The chamber was cylindrical in shape. It had two electrodes for measurement in the vacuum.The electrodes were insulated from the body of the chamber by a thick sheet of perspex. Also rubber O-rings were used to prevent vacuum leaking. It also provided freedom of the restricted movement of the electrodes (as shown in Fig [\ref{fig:electrode}]).Proper grooving was made on the electrodes (as shown in the fig [\ref{fig:chamber}]) as per our necessity. The diagram of the whole chamber is shown in fig [\ref{fig:chamber}].Several square sized copper plates (dimensions : $2\times2cm^2,3\times3cm^2,4\times4cm^2,5\times5cm^2$) were made from the workshop of physics department.These plates and the electrodes was designed in such a way that the inter plate distance can be varied in vacuum inside the chamber. \\
Next the vacuum pump assembly was designed. In light of the failure of diffusion pump in previous semester, a turbo molecular pump and a rotary pump (as backing pump) was used in the system. The number of junction points were reduced to minimize the chance of leaking. The chamber was connected to TMP via a plus type of structure shown in the fig[\ref{fig:system}]. All the sensors were mounted on one hand of the plus connector. Metal O-ring was used in between the TMP and the connector. The whole system was connected to a Kethley-6430 system and a computer to measure the data electronically without disturbing the system. The whole setup was shown in Fig [\ref{fig:system}]  \\
\\
\subparagraph{3.1.2 Experiment} \hspace{1pt} \vspace{0pt} \\
The rotary could evacuate the chamber up to 1.1 Torr. Then the TMP was used to take the vacuum up to order of $10^{-4}$ Torr. The whole experiment was done in the vacuum of this magnitude.\\
Firstly the capacitor plates were screwed with the electrodes. Proper connection was made between the computer system and the whole vacuum assembly so that whole system can be operated digitally. After evacuating the chamber to desired value, the kethley setup was switched on to start the measurement. The IV measurement was done with three different (0.5 cm, 1.0 cm, 1.5 cm ) separations between the capacitor plates.\\
The whole process of measurement was repeated for each set of copper plates. Two sets of data were taken to check the inconsistency of the measurement. The analysis of the experimental data is shown in the next section.

\begin{figure}[htbp]
	\begin{center}
		\includegraphics[width=0.3\textwidth]{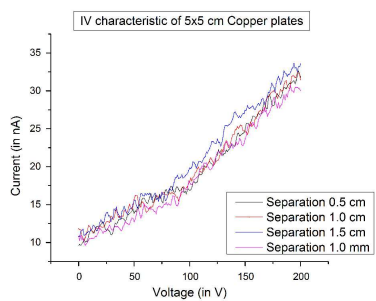}
		\caption{IV characteristic for 5x5 cm copper plates}
		\label{fig:5x5}
	\end{center}
\end{figure}

\subparagraph{3.1.3 Result \& Discussion} \hspace{1pt} \vspace{0pt} \\
The graphs obtained in the measurement are shown below.Fig [\ref{fig:2x2}],Fig [\ref{fig:3x3}], Fig [\ref{fig:4x4}],Fig [\ref{fig:5x5}].
\begin{figure}[htbp]
	\begin{center}
		\includegraphics[width=0.3\textwidth]{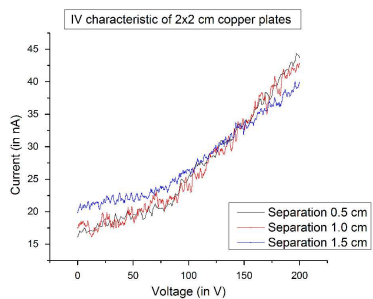}
		\caption{IV characteristic for 2x2 cm copper plates}
		\label{fig:2x2}
	\end{center}
\end{figure}\\
\begin{figure}[htbp]
	\begin{center}
		\includegraphics[width=0.3\textwidth]{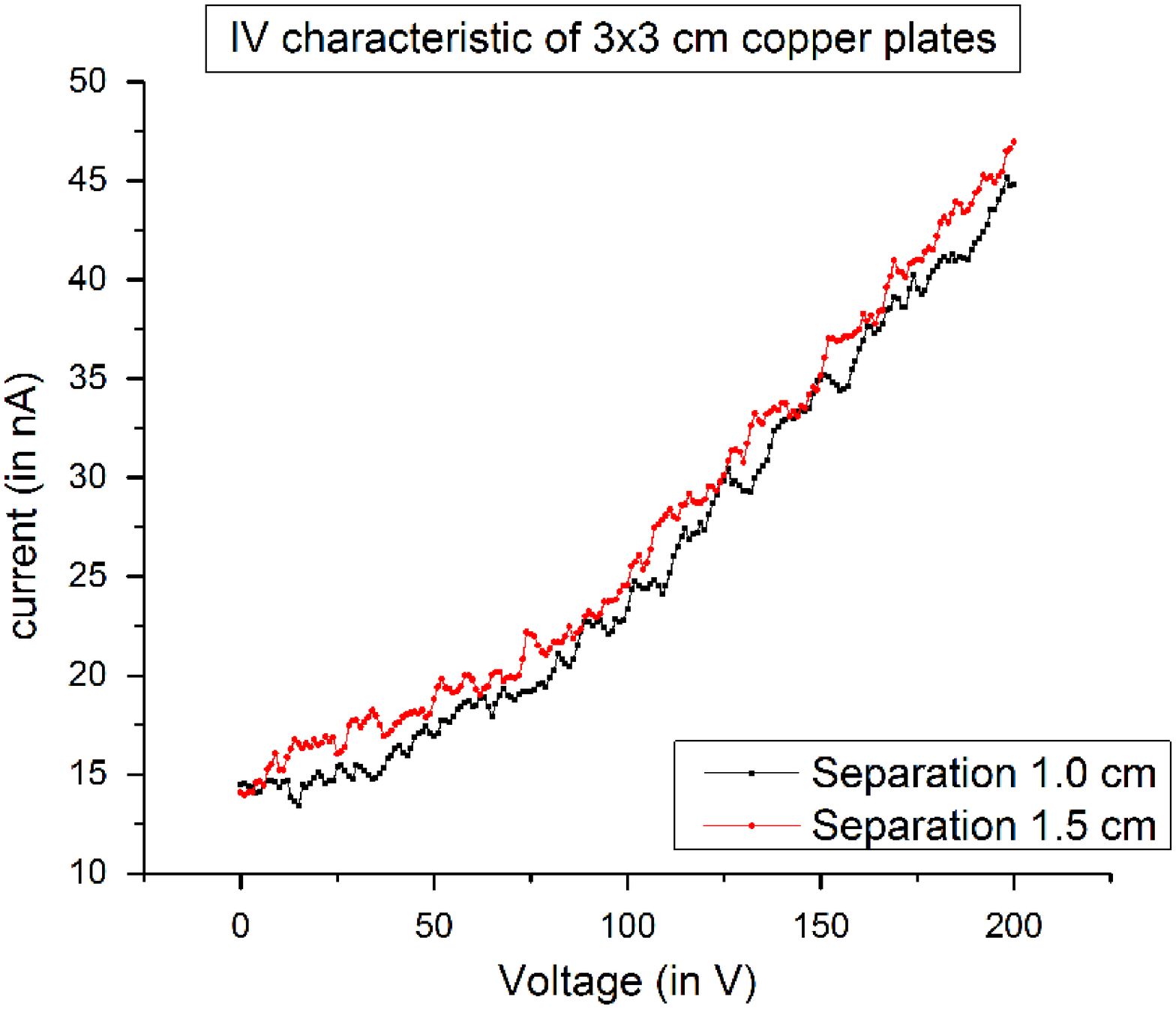}
		\caption{IV characteristic for 3x3 cm copper plates}
		\label{fig:3x3}
	\end{center}
\end{figure}\\
\begin{figure}[htbp]
	\begin{center}
		\includegraphics[width=0.3\textwidth]{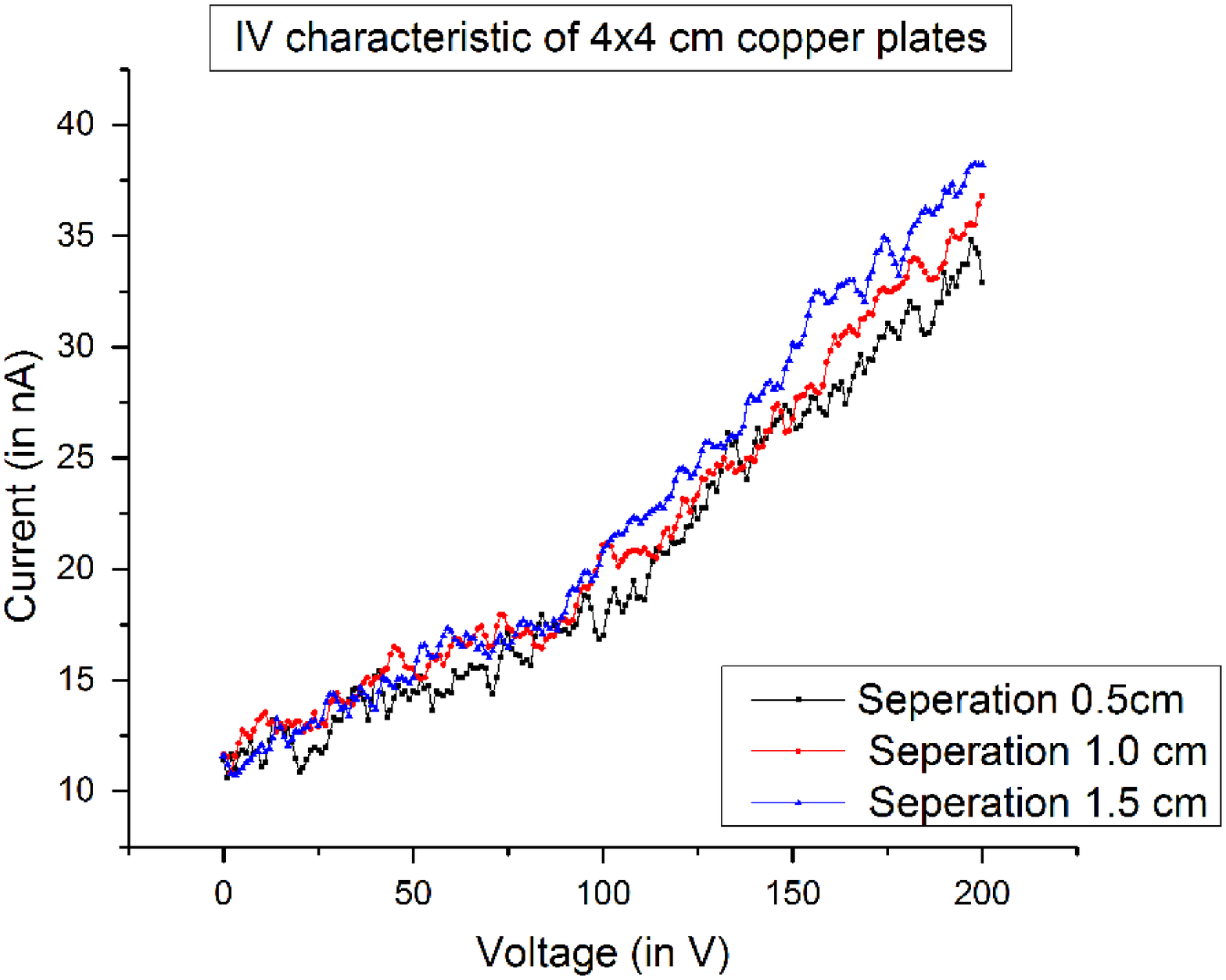}
		\caption{IV characteristic for 4x4 cm copper plates}
		\label{fig:4x4}
	\end{center}
\end{figure}\\
\\

\begin{figure}[htbp]
	\begin{center}
		\includegraphics[width=0.3\textwidth]{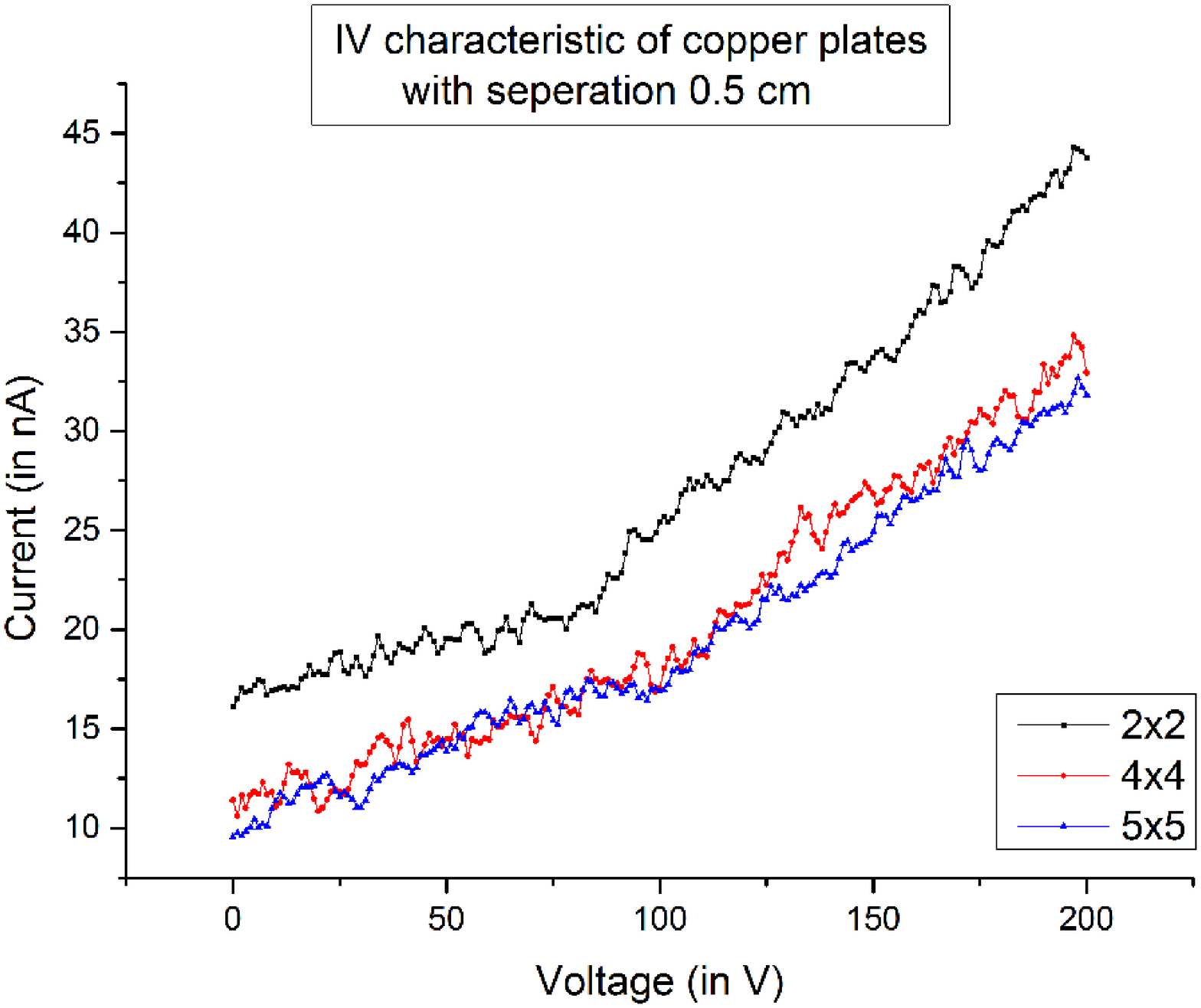}
		\caption{IV characteristic for copper plates \\(separation 0.5cm)}
		\label{fig:0.5}
	\end{center}
\end{figure}

\begin{figure}[htbp]
	\begin{center}
		\includegraphics[width=0.3\textwidth]{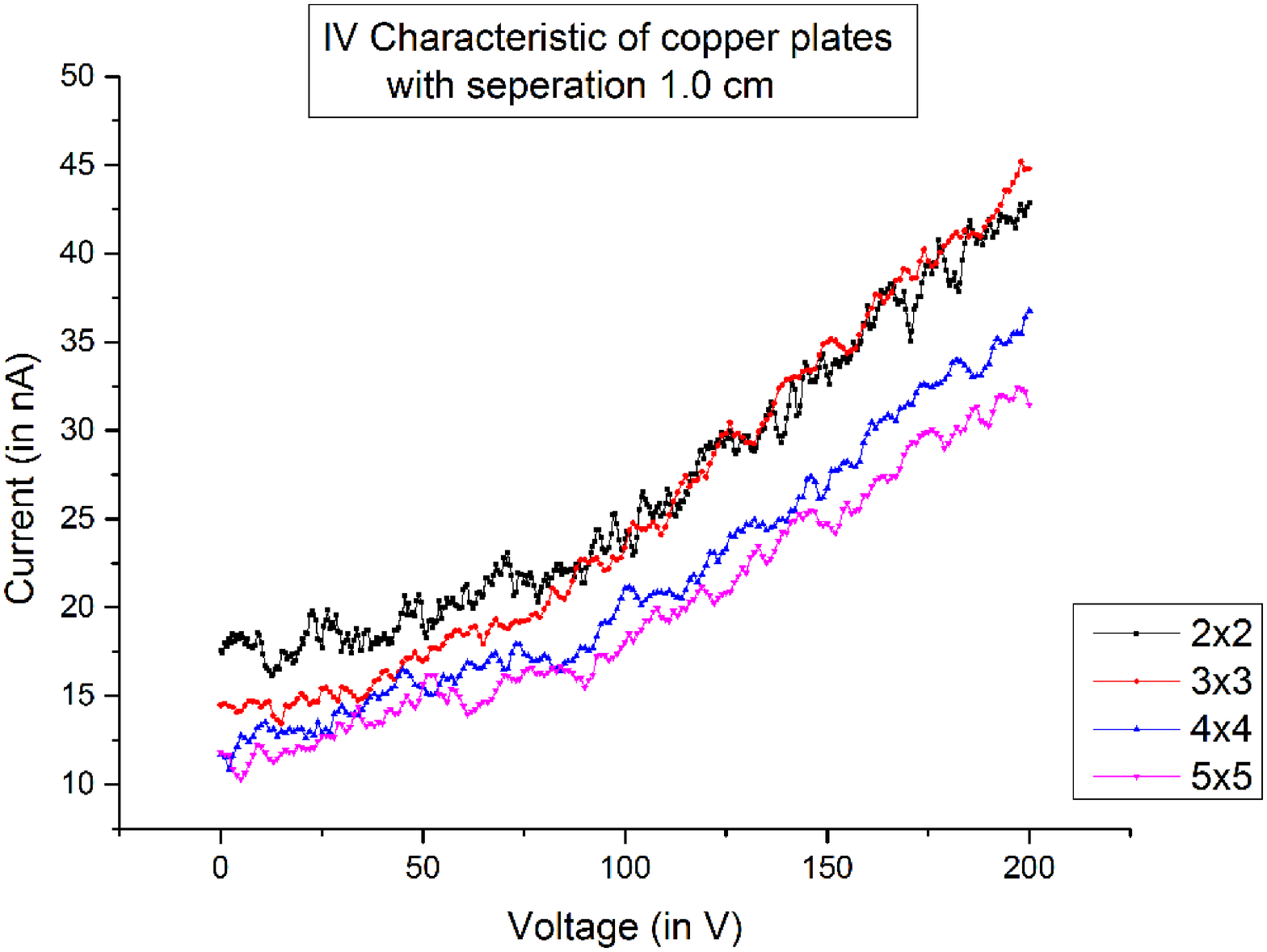}
		\caption{IV characteristic
		 for copper plates \\ (separation 1.0cm)}
		\label{fig:1}
	\end{center}
\end{figure}

\begin{figure}[htbp]
	\begin{center}
		\includegraphics[width=0.3\textwidth]{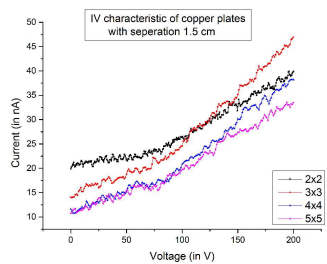}
		\caption{IV characteristic for copper plates \\ (separation 1.5cm)}
		\label{fig:1.5}
	\end{center}
\end{figure}

\begin{figure}[htbp]
	\begin{center}
		\includegraphics[width=0.3\textwidth]{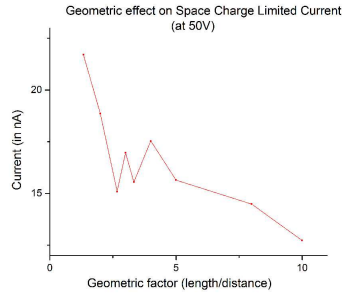}
		\caption{geometric effect on Space charge limited current\\ (at 50v)}
		\label{fig:gdep}
	\end{center}
\end{figure}

These IV characteristic curves have shown some interesting results. .These interesting results can be analyzed well when we keep separation between the plates fixed.Following graphs are plotted for different fixed values of separations.Fig[\ref{fig:0.5}],fig[\ref{fig:1}],fig[\ref{fig:1.5}]
\\
\\
\\
From these graphs one can easily notice that as indicated in the standard Child-Langmuir law, that the current varies with three and half power of the voltage, is not entirely correct. Foe example in Fig [8] ,for the same separation ,The IV curves are different for different plate dimensions.One needs to introduce a geometric factor in the expression. Such correction have been proposed in literature. Defining a geometric factor as \textbf{area/$separations^2$}.The current is supposed to be proportional to this factor.Since all other plates are of square shape,We will be using \textbf{g=length of a plate/Separation between plates} as a geometric parameter. As it is observed for small sizes the geometric correction is to be re-looked. For example, if we look at graphs in Fig [8], we see that the current largest for $2\times2cm^2$ plates and smallest for $5\times5cm^2$ for the same separation of 0.5 $cm^2$. According to geometrical correction proposed in the literature (ref [1]), the current should increase by a factor of 6.25 for a fixed voltage. In contrary, we observe the current to decrease.\\ 
With Sufficient data the geometric dependence can be obtained from these graphs. However with 4 sizes of plates and 3 separations taken one can only have 12 data points for the plotting I vs g graph as shown in fig [\ref{fig:gdep}].Two points (g=2,g=4) coincide with each other.Only Ten data points are shown in the graph. 
\\
It is very hard to obtain a power dependence from only twelve data points. However we consider geometric factor to be g and it varies to the power a.So overall we assume the current equation to be
$$
I=cg^{a}v^{b}
$$ 
where b can be 3/2 according to Child-Langmuir law or it may not be.We want to keep this freedom in the equation. From this equation,
$$
a=\frac{g}{I}\frac{dI}{dg}
$$
We have calculated slope at all the points except terminal ones and from there we get the value of a which varies roughly from -0.2 to -0.3 .The result is quite surprising.This geometric factor can be used to fit the data we obtained for the Nano capacitor measurement. However as mentioned in the introduction, if one can modify the standard Child-Langmuir law by introducing a geometric factor and if the same power holds for low dimensions them it can be immensely helpful for production of many low dimensional electronic equipments. \\

\begin{figure}[htbp]
	\begin{center}
		\includegraphics[width=0.3\textwidth]{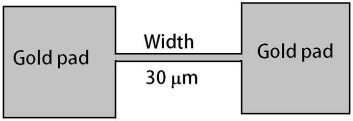}
		\caption{Mask prepared on Tin base}
		\label{fig:mask}
	\end{center}
\end{figure}

\paragraph{\normalsize{\centerline{\textbf{3.2 Studies on Nano Capacitor}}}} \hspace{1pt} \vspace{0pt} \\
\subparagraph{3.2.1 Fabrication} \hspace{1pt} \vspace{0pt} \\
In this section, The process and the designing techniques will be discussed. The Nano Capacitors was made using shadow masking technique. \\
Firstly,The mask of necessary dimensions were made at Meera lasers,a Chennai based laser drilling company. They made the mask as per our design shown in the fig [\ref{fig:mask}]on tin base. Four masks were made of width $50\mu m$,$40\mu m$,$30\mu m$,$20\mu m$ . only one mask is shown in the figure [\ref{fig:mask}].\\
\\
Then the substrate was prepared for deposition. Si wafers were oxidized to produce $SiO_2$ substrate. First, the Si wafers were cleaned by RCA technique.Then it was being heated to $1000^0$C in pure oxygen environment for oxidation for almost 24 hours.This work had been done in the Semiconductor Lab at IIT Kanpur. The $SiO_2$ was chosen as substrate because of its good adhesion property with gold film.\\
The capacitors were made by depositing gold on the $SiO_2$ substrate. Thermal Evaporator was used for the deposition. A detailed description of the thermal evaporation technique can be found in the previous semester report. The Nano wires of thickness 80 $\mu$m was prepared for the measurement.\\ 
Now Due to unavailability of the FIB machine of IBC,IITK,Somewhat crude technique was used to create the 'Nano' capacitor. A fine surgical blade was used for that purpose. A sharp cut was made by the surgical blade edge in the fine wire prepared by the shadow masking. So a small gap of the width similar to the thickness of the blade was created (as shown in the following fig[\ref{fig:setup}] ). This gap was used as the capacitor for our measurement.
\begin{figure}[htbp]
	\begin{center}
		\includegraphics[width=0.3\textwidth]{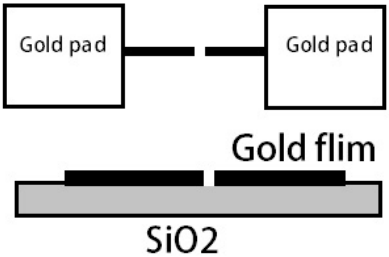}
		\caption{The nano Capacitor}
		\label{fig:setup}
	\end{center}
\end{figure}\\

\subparagraph{3.2.2 Experimentation} \hspace{1pt} \vspace{0pt} \\
A contact pin of the shape as shown in the fig [14]was used for making contact with the thin flim capacitor. This rare contact pin was provided by Prof. R.S. Anand of EE dept. of IIT kanpur. Using the pin, the capacitor was connected to a standard bread board. Standard bread board wires were used to make connection with the kethley meter.\\
\begin{figure}[htbp]
	\begin{center}
		\includegraphics[width=0.03\textwidth]{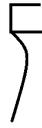}
		\caption{contact pins used}
		\label{fig:cpin}
	\end{center}
\end{figure}\\
The measurement was done on capacitors with widths 30$\mu$m and 40$\mu$m respectively. Due to technical difficulties measurement could not be done on other capacitors. Two sets of data were taken on each capacitor. The measurement was done in atmosphere.

\subparagraph{3.2.3 Result \& Discussion } \hspace{1pt} \vspace{0pt} \\
 The results are shown in the following figures.Fig[\ref{fig:30}],Fig[\ref{fig:40}]
\begin{figure}[htbp]
	\begin{center}
		\includegraphics[width=0.3\textwidth]{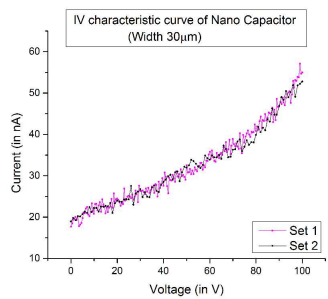}
		\caption{IV characteristic for Nano Capacitor (width 30 $\mu$m)}
		\label{fig:30}
	\end{center}
\end{figure}\\
\begin{figure}[htbp]
	\begin{center}
		\includegraphics[width=0.3\textwidth]{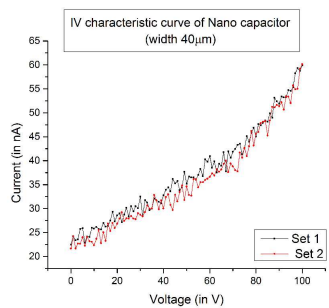}
		\caption{IV characteristic for Nano Capacitor (width 40 $\mu$m)}
		\label{fig:40}
	\end{center} 
\end{figure}\\

\subparagraph{3.2.4 Discussion} \hspace{1pt} \vspace{0pt} \\
From the graphs it can be clearly seen that it does not follow child-Langmuir law. A similar graph was shown in the previous semester report. Although in that case the gap was much larger than the present case. The similarity of the data from previous semester also indicates that they may follow the same geometric power as of macro scale study.\\
However main motive of this study is for the nano radiation detector. To distinguish the signal from the background noise we need to know the IV curve of the concerned detector without any alpha source. This data will provide the background band in the detection.Any signal which strength is stronger than the back ground can be considered as a Alpha particle hit which we will discuss below.
\section{Nano Radiation Detector}
The main idea behind the nano radiation detector is very simple. The IV characteristic between two capacitor plates are given by standard or modified Child-Langmuir law. Now this Current was generated by space charge limited current. Now if a alpha particle hits the gap in between the plates then ionization caused by the alpha particle will give rise to a lot more charge carriers. So it will result in a larger current which should be significantly distinguishable from the background signal. By proper electronics this signal can be counted and a local alpha particle distribution can be prepared which may be helpful in many scientific applications.
\subparagraph{4.1 Fabrication} \hspace{1pt} \vspace{0pt} \\
The Same capacitors used previously was used here as the detectors.
 \subparagraph{4.2 Experimentation} \hspace{1pt} \vspace{0pt} \\
 The experimental setup was also same as before. Only the Alpha particle sources was kept near the setup for radiation. Americium-241 (${}^{241}_{95}Am$ ) was used for the Alpha radiation source. All safety measures including radiation batch was taken during the experimentation.
 \subparagraph{4.3 Result \& Discussion } \hspace{1pt} \vspace{0pt} \\
 Two types of measurement were done in the experiment. Two sets of data were taken for current against voltage. Five sets of data were taken as current against time. Only one I-T graph is shown here because this is the only graph where a signal was detected.
 \begin{figure}[htbp]
 	\begin{center}
 		\includegraphics[width=0.3\textwidth]{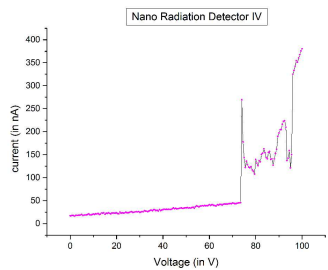}
 		\caption{IV characteristic of nano radiation detector}
 		\label{fig:nrdiv1}
 	\end{center} 
 \end{figure}\\
 \begin{figure}[htbp]
 	\begin{center}
 		\includegraphics[width=0.3\textwidth]{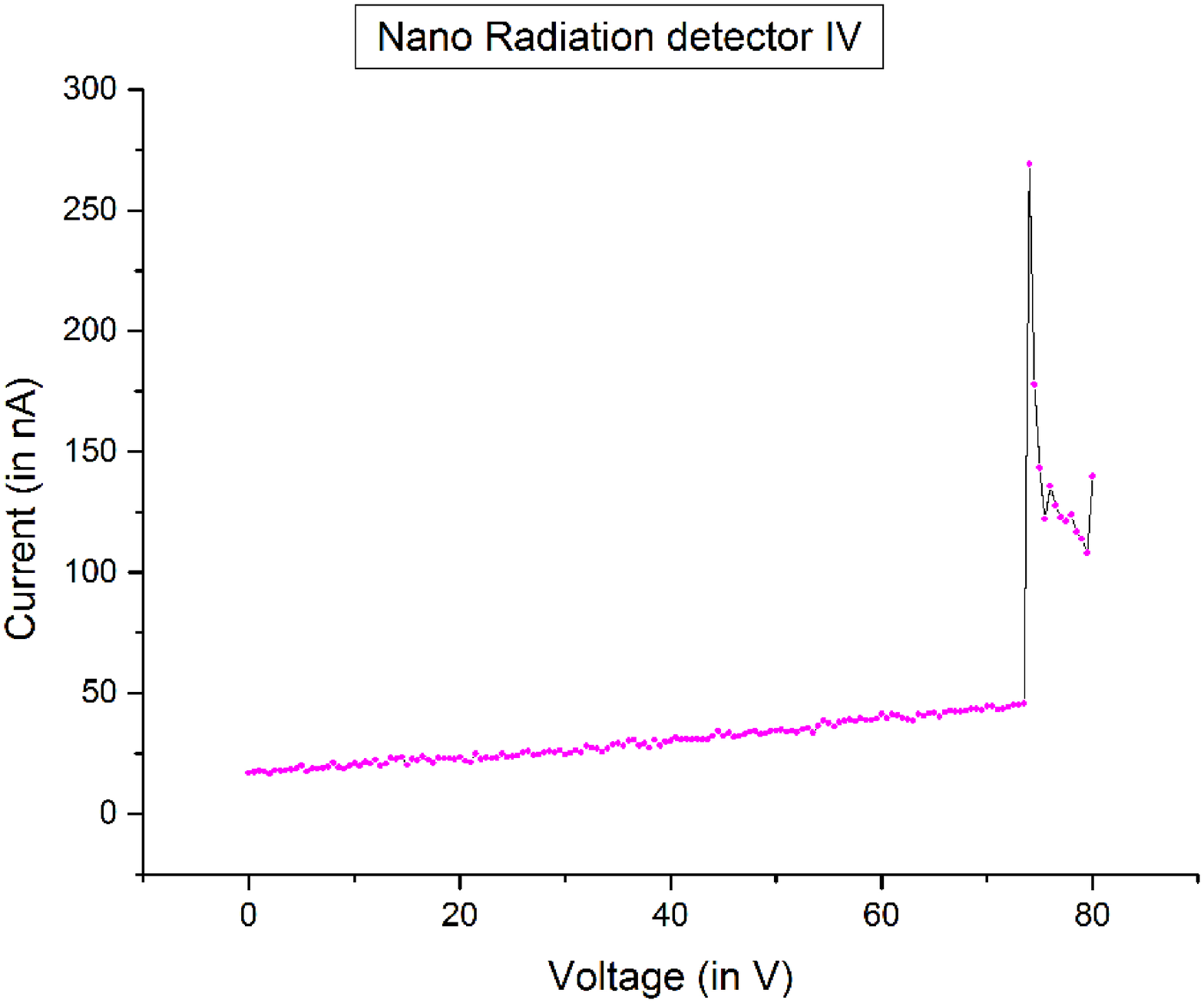}
 		\caption{IV characteristic of nano radiation detector}
 		\label{fig:nrdiv2}
 	\end{center} 
 \end{figure}\\
 \begin{figure}[htbp]
 	\begin{center}
 		\includegraphics[width=0.3\textwidth]{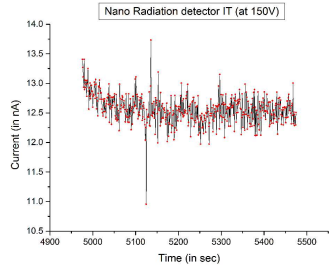}
 		\caption{IT characteristic of nano radiation detector}
 		\label{fig:nrdit}
 	\end{center} 
 \end{figure}\\
 \subparagraph{4.4 Discussion} \hspace{1pt} \vspace{0pt} \\
Now Am-241 is a strong alpha particle source. So one may expect to get a continuous hits i.e. a large number of signals. But experimental data shows opposite to the expected result. The reason behind this is the non-standardization of the equipment. Firstly the alpha source was kept just besides of the detector. So the radiation is getting spread out over a lot of area. Again we are using a single capacitor as the detector. The dimension of the wire is in micron range.Moreover the cut made on the detector also in order of microns. So it can be very easily understood that the probability of hitting event of a alpha particle at exact gap point is very low.This is the reason behind the very low hit rate. How ever this problem can be solved using a grid of capacitors which is discussed in next section.
\section{Future Agenda}
\begin{figure}[htbp]
 	\begin{center}
 		\includegraphics[width=0.2\textwidth]{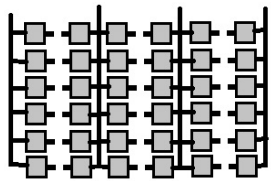}
 		\caption{Proposed grid nano radiation detector}
 		\label{fig:grid}
 	\end{center} 
 \end{figure}

From the previous discussion and the experimental result it is clear that to continue with only one capacitor for the detector is not likely to be a useful exercise. So we propose a grid of capacitors as shown below in the fig [\ref{fig:grid}] \\
\\

This type of grid will be able to detect a large number of hits of alpha particles. Moreover to ignore the capacitive effects of the pads one can use the nano capacitor study to ignore the normal background signals. The accuracy can be further increased by using the same in some special Gas environment so that the dead time can be reduced. Since the dimension of the system is very small, the dead time will be much much smaller as compared to the standard detectors.Also as we have seen in the experimental data that the signal gets stronger at higher voltages which is expected.One can use the $Ga^{+}$ irradiation data from the my colleague Avishek Kr. Basu's report to increase the voltage tolerance of nano wires of the grid.However the effect of the irradiation on the detector signal remained to be studied. We sincerely hope that some future project student will complete the work.  

\section{Acknowledgments}
I am thankful to my guide Prof H.C.Verma for his nice guidance and advises in my project.I will remain thankful to him for taking me as his student .I also express my gratitude to Prof. R.S. Anand for his help. I am also thankful to Dr. Nobin Banerji,Krishnasamy S, Sandipji for their help and support.I also thank my college avishek for his help in completing the project.Lastly i must not forget to acknowledge all my friends and professors at IIT Kanpur for their time to time valuable suggestions during discussion hours..

\end{document}